\documentclass[conference,letterpaper]{IEEEtran}
\IEEEoverridecommandlockouts

\usepackage{ifpdf}
\ifpdf
    \usepackage[pdftex]{graphicx}
    \graphicspath{{figs/}{matlab/}}   % still needed, even with Inkscape-Pdf-Latex
    \usepackage[update]{epstopdf}
\else
	\usepackage{graphicx}
	\graphicspath{{figs/}{matlab/}}   % still needed, even with Inkscape-Pdf-Latex
\fi

\usepackage[utf8]{inputenc} 
\usepackage[T1]{fontenc}
\usepackage{url}
\usepackage{ifthen}
\usepackage{amsfonts}
\usepackage{mathrsfs}
\usepackage{cite}
\usepackage[cmex10]{amsmath} % Use the [cmex10] option to ensure complicance
\usepackage{amssymb,amsthm,bm,dsfont}

\usepackage{enumerate}  
\usepackage{color}
\usepackage[hidelinks]{hyperref}
\usepackage{cleveref} % multiple refference
\usepackage{cite}
\usepackage{subcaption}
\allowdisplaybreaks[2]  %公式分页显示，数字0-4表示分开程度
\usepackage{multirow}

\usepackage{cases}
\usepackage{balance}
\newif\ifEnv

\Envtrue
\ifEnv
    
\else
    \excludecomment{Env}
\fi

\usepackage{optidef}

\newtheorem{theorem}{\textbf{Theorem}}
\newtheorem{prop}{\textbf{Proposition}}

\newtheorem{definition}{\textbf{Definition}}%[section]

\newtheorem{assumption}{\textbf{Assumption}}%[section]

\newcommand{\Expt}{\mathbb{E}}

\allowdisplaybreaks

\definecolor{bl}{rgb}{.2,.2,.8}

\usepackage{xparse}
\NewDocumentCommand{\ml}{m O{M} O{Q_n^{[M]}}} {\mathcal{L}_{#1}(#2\rightarrow #3)}

\usepackage{amsmath}
\usepackage{graphicx}

\begin{document}
\title{Source-Channel Separation Theorems for Distortion Perception Coding} 

\author{\IEEEauthorblockN{Chao Tian\IEEEauthorrefmark{1},
Jun Chen\IEEEauthorrefmark{2}, and Krishna Narayanan\IEEEauthorrefmark{3}}
\IEEEauthorblockA{\IEEEauthorrefmark{1}\IEEEauthorrefmark{3}Department of Electrical and Computer Engineering,
 Texas A\&M University, College Station, TX, 77845, USA}
\IEEEauthorblockA{\IEEEauthorrefmark{2}Department of Electrical and Computer Engineering,
 McMaster University, Hamilton, ON L8S 4L8, Canada}
{Email: \IEEEauthorrefmark{1}chao.tian@tamu.edu,
\IEEEauthorrefmark{2}chenjun@mcmaster.ca,
\IEEEauthorrefmark{3}krn@tamu.edu}}

\maketitle

\begin{abstract}
It is well known that separation between lossy source coding and channel coding is asymptotically optimal under classical additive distortion measures. Recently, coding under a new class of quality considerations, often referred to as perception or realism, has attracted significant attention due to its close connection to neural generative models and semantic communications. In this work, we revisit source-channel separation under the consideration of distortion-perception. We show that when the perception quality is measured on the block level, i.e., in the strong sense, the optimality of separation still holds when common randomness is shared between the encoder and the decoder; however, separation is no longer optimal when such common randomness is not available. In contrast, when the perception quality is the average per-symbol measure, i.e., in the weak sense, the optimality of separation holds regardless of the availability of common randomness. 
\end{abstract}

 %add counter-example for strong-sense without common randomness. 

\section{Introduction}
Shannon's source-channel separation theorem \cite{shannon1948mathematical} is a cornerstone of modern network communication systems, where layered system designs are ubiquitous. The optimality of the separation architecture continues to hold or hold approximately, even in some more complex network settings \cite{tian2010approximate,TCDS14}. However, these results were established only when the quality of the reconstruction is measured under classical additive distortion functions. 

Recently, a new type of quality consideration, often referred to as perception or realism, has emerged \cite{blau2019rethinking}. Roughly speaking, the requirement is that the reconstruction should have a probability distribution that is similar to the original source distribution. This consideration is well-motivated in neural compression, as neural generative models enjoy strong performance with visually pleasing reconstructions, which may, however, be viewed as having low quality under the traditional distortion measures such as mean squared error (MSE). Moreover, semantic communication systems may also need to be measured under such a quality consideration. 
Previous efforts in the study of coding with perception quality consideration have focused on the source coding formulation, i.e., the rate-distortion-perception function. Several conclusive results have been reported \cite{theis2021coding,zhang2021universal,chen2022rate,sadafconditional2024,hamdi2024rate}.

It should be noted that the perception quality consideration is different from the classical single-letter additive distortion measure, since the quality may be measured on the whole coding block instead of on individual sample positions. Therefore, the question of whether separation is optimal under the perception quality consideration remains mostly open. In this work, we revisit source-channel separation under the distortion-perception consideration. We show that when the perception quality is measured on the block level, i.e., in the strong sense, the optimality of separation still holds when common randomness is shared between the encoder and the decoder; however, separation is no longer optimal in general when such common randomness is not available. In contrast, when the perception quality is the average per-symbol measure, i.e., in the weak sense, the optimality of separation holds regardless of the availability of the common randomness. 

{
Our result for the case under strong-sense perception measure with common randomness is particularly notable, since the proof does not rely on a single-letter characterization of the corresponding rate-distortion-perception region. This proof allows us to establish the separation result under a general set of conditions. 
We note that in a closely related work \cite{QLCYW2024}, the problem of source-channel separations was considered when the reconstruction needs to follow a fixed target distribution in an i.i.d. manner. In contrast, in this work, we consider distortion-perception coding, where the reconstruction does not need to satisfy any specific target distribution, nor does it need to be done in an i.i.d. manner.}

The rest of the paper is organized as follows. Section \ref{sec:preliminary} introduces the notation and problem definitions. Section \ref{sec:strong} provides the main results for separation under the strong-sense measure, and Section \ref{sec:weak} gives the results for the weak-sense case. Proofs are given in Section \ref{sec:proofs}, and Section \ref{sec:conclusion} concludes the papers, with the proof for a technical proposition relegated to the appendix.

\section{Preliminaries and Problem Definitions}
\label{sec:preliminary}

\subsection{Notation and Preliminaries}
The channel we consider in this work is the classic discrete memoryless channel given by $P_{Y|X}$, and we denote its capacity as $C$. We assume the discrete memoryless source follows a probability distribution $P_S$ on a finite alphabet $\mathcal{S}$. A distortion measure function $\delta: \mathcal{S}\times \mathcal{S}\rightarrow [0,\infty)$ and a sequence of perception measure functions $d_n: \Delta_{\mathcal{S}^n}\times \Delta_{\mathcal{S}^n}\rightarrow [0,\infty]$ are assumed which satisfy $d_n(P,P)=0$, where $\Delta$ is the probability simplex on the alphabet specified by the subscript.

We will need the sequence of perception measure $d_n$ to satisfy the following regularization condition. When this condition is not satisfied, pathological cases may occur where coding over shorter sequences may have an inherent advantage.% thus the regions may not be well-defined. 

\begin{definition}
A sequence of perception measures $d_{n}: \Delta_{\mathcal{S}^{n}}\times \Delta_{\mathcal{S}^{n}}$ is called sub-decomposable, if the following condition holds for any positive integers $n_1,n_2,\ldots,n_k$ where $\sum_{i=1}^k n_i=n$ : 
\begin{align}
&d_{n}\left(\otimes_{i=1}^k P_{S_{\ell_{i-1}+1}^{\ell_i}},\otimes_{i=1}^k Q_{S_{\ell_{i-1}+1}^{\ell_i}}\right)\notag\\
&\qquad\qquad\leq \sum_{i=1}^k d_{n_i}\left(P_{S_{\ell_{i-1}+1}^{\ell_i}}, Q_{S_{\ell_{i-1}+1}^{\ell_i}}\right),
\end{align}
where $\ell_i =\sum_{j=1}^i n_j$ and $\ell_0\triangleq 0$.
\end{definition}

We state several assumptions on the perception measure below, which are not always required for our results.  
% \begin{assumption}
% \label{assumption:1}
% The normalized perception measure $\frac{1}{n}d_n(\cdot,\cdot)$ is bounded. 
% \end{assumption}

\begin{assumption}
\label{assumption:11}
The perception measure is sub-decomposable.
\end{assumption}
\begin{assumption}
\label{assumption:2}
The perception metrics $d_n$'s are continuous with respect to the total variation distance $d_{TV}$ in the sense that $d_n(P_{S^n},Q_{\hat{S}^n})\leq d_n(P_{S^n},P_{\hat{S}^n})+nc_{\max}d_{TV}(P_{\hat{S}^n},Q_{\hat{S}^n})$ for some universal constant $c_{\max}$.
\end{assumption}

\begin{assumption}
\label{assumption:3}
The perception metric $d_1$ is convex in the second argument. 
\end{assumption}

{Assumption \ref{assumption:11} allows us to establish separation results when single-letter characterizations are unavailable, and it is more lenient than the assumption in \cite{QLCYW2024}.
%Two additional technical assumptions are needed for certain results. 
%}
Several well-known measures, such as the squared Wasserstein-2 distance and the total variation distance, indeed satisfy these assumptions.% \ref{assumption:11}, \ref{assumption:2}, and \ref{assumption:3}.
}

\begin{figure}[t!]
\centering
\includegraphics[width=0.35\textwidth]{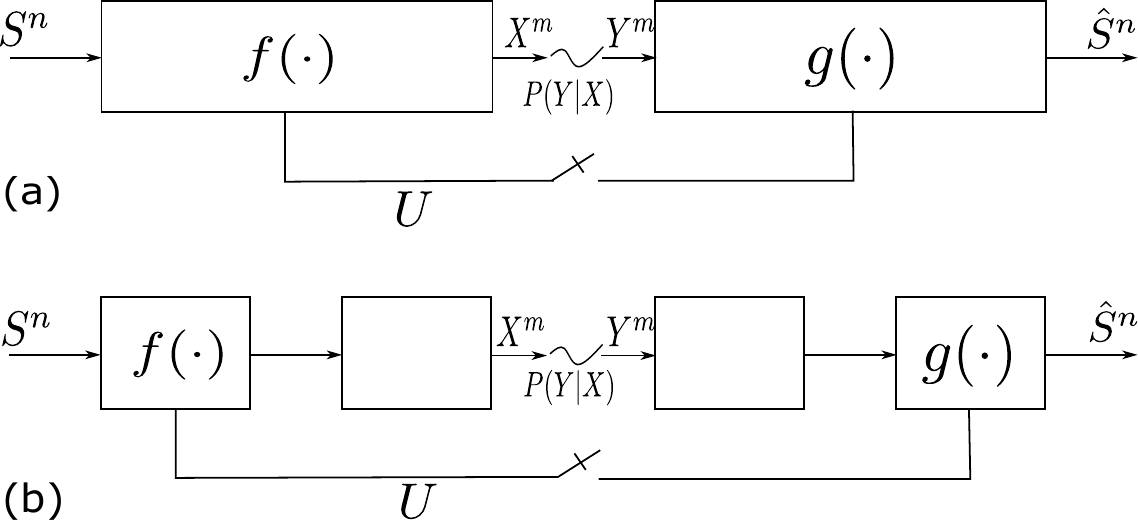}
\caption{(a) Joint source-channel coding w/ or w/o common randomness, and (b) Separate coding w/ or w/o common randomness.}
\vspace{-0.5cm}
\end{figure}

\subsection{DP Joint Source-Channel Coding Achievable Regions}

The following definition provides the encoder and the decoder for joint source-channel coding.

\begin{definition}
A distortion and perception pair $(D,P)$ is called achievable on a channel with a mismatch factor $\kappa$, if for all sufficiently large $n$, there exist common randomness $U$, an encoding function $f: \mathcal{S}^n\times \mathcal{U} \rightarrow \mathcal{X}^m$ {(where $\mathcal{U}$ is a Polish space)}, a decoding function $g: \mathcal{Y}^m\times \mathcal{U} \rightarrow \mathcal{S}^n$, such that $\hat{S}^n=g(Y^m,U)$ is the reconstruction and $Y^m$ is the channel output on the memoryless channel $P_{Y|X}$ with input vector $X^m=f(S^n,U)$, which satisfy the condition that 
\begin{align}
&\frac{m}{n}\leq \kappa,\\
&\frac{1}{n}\sum_{t=1}^n \mathbb{E}
[\delta(X_t,\hat{X}_t)]\leq D,\\
&\frac{1}{n}d_n(P_{S^n},P_{\hat{S}^n})\leq P. \label{eqn:strongsenseP}
\end{align}
The closure of the collection of all such achievable $(D,P)$ pairs is called the (strong-sense) distortion-perception joint coding achievable region, denoted as ${\mathcal{J}}$.
\end{definition}

The first condition states that the source-channel mismatch factor must be upper-bounded properly, and the second and the last conditions specify the distortion constraint and the (strong-sense) perception constraint, respectively. There are related concepts in the literature that are also of practical interest. We consider the following cases in our work:
\begin{itemize}
\item For the weak-sense perception constraint, we can simply replace (\ref{eqn:strongsenseP}) with
\begin{align}
\frac{1}{n}\sum_{t=1}^nd_1(P_{S_t},P_{\hat{S}_t})\leq P. \label{eqn:weaksenseP}
\end{align}
The corresponding achievable region is denoted as $\mathcal{J}_w$.
\item When the perception constraint is strong-sense and the randomness is not shared at the encoder and the decoder (i.e., private randomness), the problem is referred to as (strong-sense) distortion-perception joint coding without common randomness, and the corresponding achievable region is denoted as $\hat{\mathcal{J}}$.
\item When the perception constraint is weak-sense and the randomness is not shared at the encoder and the decoder, the problem is referred to as (weak-sense) distortion-perception joint coding without common randomness, and the corresponding achievable region is denoted as $\hat{\mathcal{J}}_w$.
\end{itemize}

\subsection{RDP Coding Achievable Regions}

The following definition specifies the encoder and the decoder for lossy source coding under the distortion-perception consideration. 

\begin{definition}
A rate-distortion-perception $(R,D,P)$ triple is called achievable if for all sufficiently large $n$, there exist common randomness $U$, an encoding function $f: \mathcal{S}^n\times \mathcal{U} \rightarrow \mathcal{Z}$, and a decoding function $g: \mathcal{Z}\times \mathcal{U} \rightarrow \mathcal{S}^n$, such that $\hat{S}^n=g(f(S^n,U),U)$ and $Z=f(S^n,U)$ satisfying the condition that 
\begin{align}
&\frac{1}{n}H(Z|U)\leq R,\\
&\frac{1}{n}\sum_{t=1}^n \mathbb{E}
[\delta(S_t,\hat{S}_t)]\leq D,\\
&\frac{1}{n}d_n(P_{S^n},P_{\hat{S}^n})\leq P.
\end{align}
The closure of the collection of all such achievable triples is called the (strong-sense) rate-distortion-perception achievable region, denoted as $\mathcal{G}$.
\end{definition}

{In the definition above, $Z$ can be viewed as an intermediate encoding of $S^n$ before the entropy coding.} Similar to the joint coding settings, we can define the weak-sense rate-distortion-perception achievable region, the strong-sense rate-distortion-perception achievable region without common randomness, and the weak-sense rate-distortion-perception achievable region without common randomness, respectively. We denote these regions as $\mathcal{G}_w$, $\hat{\mathcal{G}}$, $\hat{\mathcal{G}}_w$, respectively.

\begin{prop} The distortion-perception joint coding achievable region $\mathcal{J}$ and the rate-distortion-perception achievable region $\mathcal{G}$ are both convex, regardless of whether the strong-sense or the weak-sense perception measure is adopted. \label{prop:convexity}
\end{prop}

The proof is given in the appendix. As part of the proof, we also show that if a code of length $n$ exists that achieves $(D+\epsilon,P+\epsilon)$, then for any sufficiently large $n'$, there exist codes that achieve within $(D+2\epsilon,P+2\epsilon)$. Therefore, we only need to show the existence of good code for a specific $n$ instead of for all $n$ directly. 

\subsection{DP Separate Coding Regions}

We write the (strong-sense) separate coding achievable distortion-perception region as 
\begin{align}
\mathcal{J}'\triangleq\{(D,P): (R,D,P)\in \mathcal{G}, R\leq \kappa C\}.
\end{align}
We can similarly define the weak-sense separate coding achievable distortion-perception region, the strong-sense separate coding achievable distortion-perception region without common randomness, and the weak-sense separate coding achievable distortion-perception region without common randomness, denoted as $\mathcal{J}'_w$, $\hat{\mathcal{J}}'$, and $\hat{\mathcal{J}}'_w$, respectively. 

We also need the following distortion-rate-perception function in the proof:
\begin{align}
D(R,P) = \min_{(R,D,P)\in \mathcal{G}} D.
\end{align}

The regions we defined are summarized in Table \ref{tab:regions}. 
\renewcommand{\arraystretch}{1.2}
\begin{table}[t!]
    \centering
    \caption{Notations of various achievable regions, where ``cr'' means ``common randomness''}
    \label{tab:regions}
   \begin{tabular}{|c||c|c|c|} % Define 5 columns
        \hline
                   &DP joint coding & RDP & DP separate coding \\ \hline\hline
strong-sense, w/ cr     & $\mathcal{J}$ & $\mathcal{G}$ & $\mathcal{J}'$ \\ \hline
weak-sense, w/ cr       & $\mathcal{J}_w$ & $\mathcal{G}_w$ & $\mathcal{J}'_w$ \\ \hline
strong-sense, w/o cr    & $\hat{\mathcal{J}}$ & $\hat{\mathcal{G}}$ & $\hat{\mathcal{J}}'$ \\ \hline
weak-sense, w/o cr      & $\hat{\mathcal{J}}_w$ & $\hat{\mathcal{G}}_w$ & $\hat{\mathcal{J}}'_w$\\ \hline
    \end{tabular}
    \vspace{-0.4cm}
\end{table}

\section{Coding under the Strong-Sense Perception Requirement}
\label{sec:strong}

\subsection{Separation: Common Randomness Present}

Under the strong-sense perception requirement, when there exists common randomness shared by the encoder and the decoder, source-channel separation is optimal in general. We state the precise results in the following theorems. 
\begin{theorem}
\label{thoerem:converse}
%Under Assumption \ref{assumption:1} and \ref{assumption:11}, 
$\mathcal{J}'\supseteq\mathcal{J}$.
\end{theorem}

\begin{theorem}
\label{theorem:achievable}
Under Assumptions \ref{assumption:11} and \ref{assumption:2}, we have $\mathcal{J}'\subseteq\mathcal{J}$. 
\end{theorem}

As a consequence, under Assumptions \ref{assumption:11} and \ref{assumption:2}, we have $\mathcal{J}'=\mathcal{J}$, which essentially says the separate coding achievable region is exactly the same as the joint coding achievable region under these conditions. We remark that a single-letter characterization of the achievable rate-distortion-perception region is not known under Assumptions \ref{assumption:11} and \ref{assumption:2}. The single-letter results given in \cite{chen2022rate} were in fact established under a more stringent requirement on the perception functions. Nevertheless, we are still able to establish the optimality of the separation architecture using a proof that does not rely on the single-letter characterization.

\subsection{Failure of Separation: Common Randomness Absent}

Under the strong-sense perception requirement, when there is no common randomness shared by the encoder and the decoder, source-channel separation is not optimal in general. We next provide a counter-example from \cite{QLCYW2024} to illustrate this. 

Let $P_S$ be a binary uniform distribution and $P_{Y|X}$  a binary-symmetric channel with crossover probability $p\in(0,\frac{1}{2})$. Consider the special case where $\kappa=1$. It is evident that the uncoded scheme achieves strong-sense perfect realism (i.e., $P=0$) with an end-to-end average Hamming distortion of $p$.  In contrast, according to \cite[Example 1]{chen2022rate}, the minimum achievable distortion with separate coding under the perfect realism constraint is $2p(1-p)$, which is strictly greater than $p$. This strict separation implies that separation is not optimal.

\section{Coding under the Weak-Sense Perception Requirement}
\label{sec:weak}

The problem under the weak-sense perception measure is considerably simpler: Separation is optimal in general, and moreover, there is no essential benefit for having common randomness over the lack of it. We have the following theorem. 
\begin{theorem}
\label{theorem:weak}
{Under Assumptions \ref{assumption:2} (which only needs to hold with $n=1$) and \ref{assumption:3},} 
$\mathcal{J}_w=\mathcal{J}'_w=\hat{\mathcal{J}}_w=\hat{\mathcal{J}}'_w=\mathcal{J}^*_w$, where 
\begin{align}
\mathcal{J}^*_w\triangleq &\bigg{\{}(D,P): I(S;\hat{S})\leq \kappa C,\, \Expt \delta(S,\hat{S})\leq D, \notag\\
&\qquad\qquad d_1(S,\hat{S})\leq P, \text{ for some } P_{\hat{S}|S}\bigg{\}}.
\end{align}
\end{theorem}

\section{Proofs of Main Theorems}
\label{sec:proofs}

\subsection{Proof of Theorem \ref{thoerem:converse}}
\begin{proof}
Consider any $(D,P)$ pair that is achievable with the channel mismatch parameter $\kappa$, and choose a sufficiently large $n$ such that a code specified by $(f,g)$, with common randomness $U$ that is independent of $S^n$, can achieve the pair $(D+\epsilon,P+\epsilon)$. Note the joint distribution in this joint code can be factorized as \begin{align}
P_{S^n}P_{U}P_{X^m|S^n,U}P_{Y^m|X^m}P_{\hat{S}^n|Y^m,U}.\label{eqn:Pjoint}
\end{align}
We wish to show that there exists a rate-distortion-perception code that achieves with $R\leq \kappa C+\delta$ for any $\delta>0$. By the strong functional representation lemma \cite{li2018strong}, there exists a random variable $U'$, independent of $S^n$ such that $\hat{S}^n$ can be expressed as a deterministic function of $(S^n,U')$ and 
\begin{align}
H(\hat{S}^n|U')\leq I(S^n;\hat{S}^n)+\log(I(S^n;\hat{S}^n)+1)+4.
\end{align}
Setting $Z=\hat{S}^n$, we get
\begin{align}
\frac{1}{n}H(Z|U')&\leq \frac{1}{n}I(S^n;\hat{S}^n)+\frac{1}{n}\log(I(S^n;\hat{S}^n)+1)+\frac{4}{n}\nonumber\\
&\leq \frac{m}{n}C+\frac{1}{n}\log(mC+1)+\frac{4}{n},
\end{align}
where we utilized %$I(X^m;Y^m)\leq mC$ due to the classical channel coding theorem. 
the fact that
\begin{align}
I(S^n;\hat{S}^n)&\leq I(S^n;\hat{S}^n,U)=I(S^n;\hat{S}^n|U)\notag\\
&\leq I(X^m;Y^m|U)\leq I(U,X^m;Y^m)\notag\\
&=I(X^m;Y^m)\leq mC,\label{eqn:converse}
\end{align}
where the inequality $I(S^n;\hat{S}^n|U)\leq I(X^m;Y^m|U)$ holds because for each fixed $U=u$, data processing inequality implies $I(S^n;\hat{S}^n|U=u)\leq I(X^m;Y^m|U=u)$, and the last equality is due to the Markov string $U\leftrightarrow X^m\leftrightarrow Y^m$ implied by the factorization given above. It follows that
\begin{align}
&\frac{1}{n}H(Z|U')<\kappa C +\delta_{n,m},
\end{align}
for some positive $\delta_{n,m}$ which diminishes as $n,m$ become large. Since our new code preserves $p_{S^n,\hat{S}^n}$, we have
\begin{align}
&\frac{1}{n}\sum_{t=1}^{n}\Expt \delta(S_t,\hat{S}_t)\leq D+\epsilon,\\
& \frac{1}{n}d_{n}(P_{S^{n}},P_{\hat{S}^{n}})\leq P+\epsilon.
\end{align}
The non-negative quantity $\epsilon,\delta_{m,n}$ can be made arbitrarily small by driving $n,m$ to infinity. Therefore we indeed have $(\kappa C,D,P)\in \mathcal{G}$. 
This shows that $(D,P)\in \mathcal{J}'$, therefore $\mathcal{J}\subseteq \mathcal{J}'$. The proof is complete. 
\end{proof}

\subsection{Proof of Theorem \ref{theorem:achievable}}

\begin{proof}
{
It suffices to prove that
\begin{align}
\mathcal{J}\supseteq\{(D,P): (R,D,P)\in \mathcal{G}, R<\kappa C\},
\end{align}
instead of $\mathcal{J}\supseteq \mathcal{J}'$ directly, because the joint coding achievable distortion-perception region $\mathcal{J}$ is closed by definition, and $D(R,P)$ is convex for any fixed value $P\geq 0$, and thus continuous in the relative interior. }

%as the rate-distortion-perception function is continuous. The continuity is implied by the fact that the joint coding achievable distortion-perception region is closed by definition, and the rate-distortion-perception function is continuous in the relative interior of the non-negative quadrant (implied by its convexity). 

Let $(R,D,P)\in \mathcal{G}$ and $R<\kappa C$, there exists a source code $(f,g)$ satisfying Definition 3 with a sufficiently large $n$, i.e.,
\begin{align}
&\frac{1}{n}H(Z|U)\leq R+\epsilon\\
&\frac{1}{n}\sum_{t=1}^{n} \mathbb{E}
\delta(S_t,\hat{S}_t)]\leq D+\epsilon\\
&\frac{1}{n}d_n(P_{S^{n}},P_{\hat{S}^{n}})\leq P+\epsilon.
\end{align}
With this code, let us consider its length-$k$ extension, with $U_i$'s mutually independent for $i=1,2,\ldots,k$, {which would allow us to address the event when the description $Z$ (conditioned on $U$) is too long to fit in a channel coding block.} Let $P_{S^{nk},\hat{S}^{nk}}$ denote the joint distribution of $(S^{nk},\hat{S}^{nk})$ induced by this extended code. 

When $k$ is large, we have $\frac{1}{nk}\sum_{i=1}^k H(Z_i|U_i)\leq R+\delta$ holds with high probability. More precisely, choose an arbitrary pair $\delta_R,\epsilon_R> 0$, there exists a sufficiently large $k$ such that
\begin{align}
&Pr(\text{the sum of description lengths of $Z_1,\ldots,Z_k$}\notag\\
&\qquad\qquad\qquad\qquad\geq nk(R+\epsilon+\delta_R))\leq \epsilon_R.
\end{align}
We shall send the cumulative descriptions of the $k$-extension of the original rate-distortion perception code, if the cummulative description length is less than $nk(R+\epsilon+\delta_R)$, using an almost capacity-achieving channel code; otherwise, we simply send a fixed dummy sequence of a shorter length (which will be viewed as an error event correspondingly).
%The rate is therefore upper bounded by $R+\epsilon+\delta_R+\frac{1}{nk}$ in both cases. 

{With $\delta_R, \epsilon$ chosen to be any fixed small quantities,} since $R<\kappa C$, when $n,k$ are both sufficiently large, there exists a channel code at rate greater than $\kappa^{-1}(R+\epsilon+\delta_R+\frac{1}{nk})$, such that the decoding error probability is upper-bounded by $\epsilon_C$, where $\epsilon_C$ diminishes as $n,k$ are sufficiently large. The new decoder first attempts to recover the intended $Z^k$ sequences, and let $\hat{Z}^k$ denote this reconstruction. 
Two error events (when $Z^k\neq \hat{Z}^k$) may occur: 1) The description of the concatenated source sequence $Z^k$ is too long and the fixed dummy sequence was sent, and 2) A channel decoder error occurred. By the union bound, we have 
\begin{align}
Pr(Z^k\neq\hat{Z}^k)\leq\epsilon_R+\epsilon_C.
\end{align}
The new joint decoder will invoke the original decoding function $g$ on each individual $Z_i$ and the corresponding common randomness $U_i$, and concatenate these outputs to obtain $\hat{S}^{nk}$. When $Z^k=\hat{Z}^k$, i.e., $\hat{Z}^k$ was correctly recovered, this would produce the same output as the original source code $(f,g)$; in the event $Z^k\neq\hat{Z}^k$, an arbitrary reconstruction sequence would be obtained. Denote the joint distribution between ${S}^{nk}$ and $\hat{S}^{nk}$ using this new code over the given channel as $Q_{S^{nk},\hat{S}^{nk}}$, it follows by \cite[Lemma 2]{song2016likelihood} that
\begin{align}
d_{TV}(P_{S^{nk},\hat{S}^{nk}},Q_{S^{nk},\hat{S}^{nk}})\leq Pr(Z^k\neq\hat{Z}^k)\leq\epsilon_R+\epsilon_C,
\end{align}
Therefore, 
\begin{align}
%&\frac{1}{nk}H(Z^k|U^k)\leq R+\delta+\\
&\frac{1}{nk}\sum_{t=1}^{nk} \mathbb{E}_Q[
\delta(S_t,\hat{S}_t)]\notag\\
&\leq\frac{1}{nk}\sum_{t=1}^{nk} \mathbb{E}_P[
\delta(S_t,\hat{S}_t)]+2(\epsilon_R+\epsilon_C)D_{\max}\notag\\
&\leq D+\epsilon+2(\epsilon_R+\epsilon_C) D_{\max},
\end{align}
where we have used eqn. (29) in \cite{cuff2013distributed}, 
and 
\begin{align}
&\frac{1}{nk}d_{nk}(P_{S^{nk}},Q_{\hat{S}^{nk}})\notag\\
&\leq \frac{1}{nk}d_{nk}(P_{S^{nk}},P_{\hat{S}^{nk}})+(\epsilon_R+\epsilon_C)c_{\max}\notag\\
&\leq P+\epsilon+(\epsilon_R+\epsilon_C)c_{\max},\label{eq:sub-decomposability&continuity}
\end{align}
where \eqref{eq:sub-decomposability&continuity}  is due to the fact that $d_{nk}$ is continuous with respect to $d_{TV}$ (i.e., Assumption \ref{assumption:2}) and sub-decomposable. The code clearly satisfies the channel mismatch factor $\kappa$.
\end{proof}

\subsection{Proof of Theorem \ref{theorem:weak}}

\begin{proof}

In the weak-sense setting, it is clear by definition that $\mathcal{J}_w\supseteq \hat{\mathcal{J}}_w$,   $\mathcal{J}_w\supseteq \mathcal{J}'_w$, and $\hat{\mathcal{J}}_w\supseteq \hat{\mathcal{J}}'_w$. {It was also shown in \cite{chen2022rate} that under Assumption \ref{assumption:2} (which only needs to hold for $n=1$) and Assumption \ref{assumption:3}, the weak-sense achievable rate-distortion-perception region is characterized as} 
\begin{align}
\mathcal{Q}_w=\hat{\mathcal{Q}}_w=\mathcal{Q}^*_w,
\end{align}
where 
\begin{align}
\mathcal{Q}^*_w \triangleq &\bigg{\{}(R,D,P): I(S;\hat{S})\leq R,\, \Expt \delta(S,\hat{S})\leq D, \notag\\
&\qquad\qquad d_1(S,\hat{S})\leq P, \text{ for some } P_{\hat{S}|S}\bigg{\}}.
\end{align}
It follows that $\mathcal{J}'_w=\hat{\mathcal{J}}'_w$. Therefore, we only need to establish 
$\mathcal{J}_w\subseteq \mathcal{J}^*_w$ and $\hat{\mathcal{J}}'_w\supseteq \mathcal{J}^*_w$ to complete the proof.  

To show $\mathcal{J}_w\subseteq \mathcal{J}^*_w$, consider an arbitrary pair $(D,P)\in \mathcal{J}_w$. For any $\epsilon>0$, there exists a perception-distortion code $(f,g)$ that achieves $(D+\epsilon,P+\epsilon)$. For this code, we have 
\begin{align}
mC&\geq I(S^n;\hat{S}^n)=\sum I(S_t;\hat{S}^n|S_1^{t-1})\notag\\
&\geq \sum I(S_t;\hat{S}^n)\geq \sum I(S_t;\hat{S}_t)=nI(S;\hat{S}),
\end{align}
where the first inequality is by the same steps as in (\ref{eqn:converse}), and in the last step we define $\hat{S}=S_T$ with $T$ being a random variable uniformly distributed in the set $\{1,2,\ldots,N\}$. It follows that
\begin{align}
\kappa C\geq \frac{m}{n}C\geq I(S;\hat{S}).
\end{align}
We also have
\begin{align}
&D+\epsilon\geq \frac{1}{n}\sum_t \Expt \delta(S_t,\hat{S}_t)=\Expt \delta(S,\hat{S})\\
&P+\epsilon\geq \frac{1}{n}\sum_t \Expt d_1(S_t,\hat{S}_t)\geq\Expt d_1(S,\hat{S}),\label{eqn:Pepsilon}
\end{align}
where Assumption \ref{assumption:3} is used in (\ref{eqn:Pepsilon}). Since $\epsilon$ can be made arbitrarily small by making $n$ large, we have $(D,P)\in \mathcal{J}^*_w$.

To show $\hat{\mathcal{J}}'_w\supseteq \mathcal{J}^*_w$, we shall rely on the combination of the coding scheme given in \cite{chen2022rate} for the rate-distortion-perception problem and the conventional capacity-achieve channel codes.  Similarly to the proof of Theorem \ref{theorem:achievable}, the only issue that needs to be resolved is that due to excess source coding description length and channel uncertainty, the description $Z$ has a diminishing error probability of being decoded erroneously at the decoder. However, the penalties such induced on the distortion and the perception are both negligible, when Assumption \ref{assumption:2} holds with $n=1$. The proof is thus complete.
\end{proof}

\section{Conclusion}
\label{sec:conclusion}

We consider source-channel separation with perception consideration in addition to the standard distortion measure. We show that under the strong-sense perception constraint, separation is optimal when common randomness is available at the encoder and the decoder; however, when common randomness is not available, source-channel separation is not optimal in general. In contrast, under the weak-sense perception constraint, separation is always optimal regardless of the existence of common randomness.  

{\section*{Acknowledgement} 
The authors are grateful to the anonymous reviewer who provided numerous constructive comments, which helped to improve the quality of this work. }

%\newpage
\appendices
\section{Proof of Proposition \ref{prop:convexity}}
\begin{proof}
This convexity can be established using a time-sharing argument. We give a brief outline of this argument for the distortion-perception joint coding achievable region $\mathcal{J}$ under the strong-sense definition. The proofs for the other cases are similar. 

Consider two achievable distortion-perception pairs $(D_1,P_1)$ and $(D_2,P_2)$, for sufficiently large $n$ there exist codes $(f_1,g_1)$ and $(f_2,g_2)$ that achieve within $\epsilon$ of the distortion-perception pairs, each taking $m$ channel uses where $m\leq \kappa n$. We wish to show that the distortion perception pair $(\lambda D_1+(1-\lambda)D_2,\lambda P_1+(1-\lambda)P_2)$ is also achievable for any $\lambda\in(0,1)$. For any $\delta>0$, there exists a sufficient large integer $k_1,k$, such that $|k_1/k-\lambda|\leq \delta$. We shall concatenate $k_1$ independent copies of $(f_1,g_1)$ and $k-k_1$ independent copies of $(f_2,g_2)$. This new code of length-$kn$ uses the channel $km$ times. The induced distortion is therefore upper-bounded by
\begin{align}
&\frac{k_1}{k}(D_1+\epsilon)+\frac{k-k_1}{k}(D_2+\epsilon)\notag\\
&\qquad\leq \lambda D_1+(1-\lambda)D_2 +\epsilon + 2\delta D_{\max},
\end{align}
where $D_{\max}$ is the upper bound of the distortion.
Similarly, for the perception
\begin{align}
&\frac{1}{kn}d_{kn}(P_{S^{kn}},P_{\hat{S}^{kn}})\leq 
\frac{k_1}{k}(P_1+\epsilon)+\frac{k-k_1}{k}(P_2+\epsilon)\notag\\
&\qquad\qquad\leq \lambda P_1+(1-\lambda)P_2 +\epsilon + 2\delta \max(P_1,P_2),\label{eqn:boundP}
\end{align}
where we utilize the sub-decomposability of the perception measures in the first inequality. By making $k$ and $n$ both large, $\epsilon$ and $\delta$ can both be made arbitrarily small. Therefore, there exists a code of length $n'=kn$ that achieves within $\epsilon'$ of the target $(D,P)$ pair. 

We then show that as long as there exists a code function pair ($f,g)$ of length-$n$ that achieves within $\epsilon$ of the constraint $(D,P)$, codes exist for all sufficiently large $n'$ to be within $2\epsilon$ of the constraint $(D,P)$. This can again be obtained by concatenating $k$ code copies of length-$n$ with a sufficiently large $k$, {and some random segment (with zero perception loss) of length-$n''$ to obtain a code of length-$(kn+n'')$}, where $0\leq n''<n$. When $k$ is sufficiently large, since the distortion function is bounded and the perception measure function is sub-decomposable, the segment of length-$n''$ only induces {a negligible difference in the eventual distortion (and zero perception loss)}. This completes the proof. 
\end{proof}

% \textcolor{red}{
% The only passage in the proofs for which it is important that the authors make modifications/clarifications is the continuity argument at the beginning of the proof of Theorem 2. I believe that the argument would be valid if one invokes the continuity of $(R,D) \mapsto min_{P s.t. (R,D,P) \in G} P$ -or the same which D and P switched-, but that the argument is not valid if only the continuity of $(D,P) \mapsto min_{R s.t. (R,D,P) \in G} R$ is invoked. Moreover, Eq. (19) and the sentence it is included in seem to be incomplete. Perhaps it is necessary to add “ J’ = ” in Eq. (19).}

% \textcolor{red}{
% I would like to point out that I believe that Assumption 1 - which limits the set of perception measures one can use - can be relaxed to "$d_n$ does not take infinite values". Indeed, in the proof of Proposition 1, the perception performance of the random length-n'' code would still be necessarily bounded -under the relaxed assumption- as k goes to infinity because n'' can only take a finite number of values and the alphabets are finite.} % Moreover, in the same proof, P1 and P2 can simply be upper bounded by $max(P1,P2)$.}

% \textcolor{red}{
% I would also like to point out that I believe that Theorem 1 holds without assuming Assumptions 1 and 2, because G and J are defined as closures. Moreover, there is no use of convexity in the proof of Theorem 1, nor of the property that it is sufficient to find one n such that there exists a good code.
% }

\textcolor{red}{
}
\bibliographystyle{IEEEtran}
\bibliography{DP}
\end{document}